\documentclass[9pt,conference]{IEEEtran}
\IEEEoverridecommandlockouts

\newif\ifANONYMOUS
\ANONYMOUSfalse    % <-- for camera-ready

\usepackage{subfiles}
\usepackage{soul}
\usepackage{xcolor}
\usepackage{graphicx}
\usepackage{svg}
\graphicspath{{./images/}}
\usepackage{cite}

\usepackage{listings}
\newcommand{\llm}{GPT-4o}

\usepackage[export]{adjustbox} % Aligning differently-sized subfigures
\usepackage{algorithmic}
\usepackage{amsmath}
\usepackage{amsfonts}
\usepackage{array}
\usepackage[USenglish]{babel}
\usepackage{balance}
\usepackage{booktabs}
\usepackage{colortbl}
\usepackage{hyperref} % Must be loaded before cleveref
\usepackage{cleveref}
\usepackage{enumerate}
\usepackage{float}
\usepackage{forest}
\usepackage[T1]{fontenc}
\usepackage{graphicx}
\usepackage[htt]{hyphenat}
\usepackage[utf8]{inputenc}
\usepackage{csquotes} % Must be loaded after inputenc
\usepackage{makecell}
\usepackage{multirow}
\usepackage{textgreek}
\usepackage{soul}
\usepackage{paralist}
\usepackage{pifont}
\usepackage{subcaption} % Subfigures etc.
\usepackage{tcolorbox}
\tcbuselibrary{breakable}
\usepackage{textcomp}
\usepackage{xurl}
\usepackage{xcolor}
\usepackage{xspace}

\usepackage{tikz} % Import the tikz package
\usetikzlibrary{automata} % Import library for drawing automata
\usetikzlibrary{positioning} % ...positioning nodes
\usetikzlibrary{arrows} % ...customizing arrows
\usetikzlibrary{calc} % ...fancy edge calculations
\usetikzlibrary{patterns} % ...shading
\usetikzlibrary{snakes} % ...snakes
\usetikzlibrary{shadows} % ...shadows
\usetikzlibrary{shapes} % ...shapes
\usetikzlibrary{arrows.meta} % ...arrows

\usepackage{mdframed}
\mdfsetup{skipabove=0.5\topskip,skipbelow=0.5\topskip,align=center}

\usepackage{amsthm}
\newtheorem{thm}{Theorem}%[section]
\usepackage{enumitem}
\setlist[itemize]{leftmargin=*,noitemsep,topsep=0pt}
\setlist[enumerate]{leftmargin=*}

\newlist{researchquestions}{enumerate}{1}
\setlist[researchquestions]{label*=\textbf{RQ\arabic*}, leftmargin=*}

\crefformat{section}{\S#2#1#3}
\crefmultiformat{section}{\S#2#1#3}{, \S#2#1#3}{, and \S#2#1#3}{}
\crefname{figure}{Figure}{Figures}
\crefname{appendix}{Appendix}{Appendices}
\crefname{table}{Table}{Tables}
\crefname{algorithm}{Algorithm}{Algorithms}
\crefname{listing}{Listing}{Listings}
\crefname{theorem}{Theorem}{Theorems}
\crefname{thm}{Theorem}{Theorems}
\crefname{lemma}{Lemma}{Lemmata}
\crefname{equation}{Eqt.}{Eqts.}
\crefformat{Grammar}{Grammar #1}

\PassOptionsToPackage{final}{commenting}

\usepackage[nompar]{misc/commenting}

\declareauthor{ET}{Eray}{red}
\declareauthor{BC}{Berk}{magenta}
\declareauthor{AK}{Atakan}{blue}
\declareauthor{VH}{Vahid}{pink}

\authorcommand{ET}{comment}
\authorcommand{BC}{comment}
\authorcommand{AK}{comment}
\authorcommand{VH}{comment}

\onlyauthors{ET,BC,AK,VH}

\newcommand{\TODO}[1]{\todo{\textbf{TODO}\\{#1}}}

\begin{document}
\newcommand{\paragraphTitle}[1]{\textbf{\textit{#1.}}}

\title{PR-Aware Automated Unit Test Generation: Challenges and Opportunities}

\ifANONYMOUS
  \author{
    \IEEEauthorblockN{Anonymous Authors}
    \IEEEauthorblockA{
      % Institution info removed for double-blind review
    }
  }
\else
  \author{
    \IEEEauthorblockN{Vahid Haratian}
    \IEEEauthorblockA{
      \textit{Bilkent University}\\
      Ankara, Turkey\\
      vahid.haratian@bilkent.edu.tr
    }
    \and
    \IEEEauthorblockN{Atakan Akar}
    \IEEEauthorblockA{
      \textit{Bilkent University}\\
      Ankara, Turkey \\
      atakan.akar@ug.bilkent.edu.tr
    }
    \and
    \IEEEauthorblockN{Berk Çakar\textsuperscript{$\dagger$}}
    \IEEEauthorblockA{
      \textit{Purdue University}\\
      West Lafayette, IN, USA \\
      bcakar@purdue.edu
    }
    \and
    \IEEEauthorblockN{Eray Tüzün}
    \IEEEauthorblockA{
      \textit{Bilkent University}\\
      Ankara, Turkey \\
      eraytuzun@cs.bilkent.edu.tr
    }
  }
\fi

\maketitle

\ifANONYMOUS
\else
  \begingroup
  \renewcommand\thefootnote{$\dagger$}
  \footnotetext{The author’s contributions to this work were made while he was an undergraduate student at Bilkent University.}
  \endgroup
\fi

\begin{abstract}
Automated test generation has a substantial body of work, yet most studies focus on generating tests for complete software units (e.g., classes) and rely on metrics such as code coverage for assessment. 
In contrast, modern software development primarily evolves through small, targeted changes introduced in pull requests (PRs). 
Despite this, the crucial task of generating tests specifically for these PRs has been overlooked, and the performance of state-of-the-art tools for this purpose remains unknown.

This study evaluates two distinct approaches for PR-aware test generation: EvoSuite, a leading search-based tool, and GPT-4o, one of the widely used large language models (LLM). 
To measure their effectiveness at validating PR-specific changes, we assess their ability to generate "fail-to-pass" (F2P) test cases—tests that fail on the code before the change and pass on the code after the change.

Our evaluation shows that EvoSuite outperformed GPT-4o, producing at least one F2P test for a significantly higher percentage of PRs (36\% vs. 13\%)
The performance of GPT-4o was significantly hampered by a high rate of compilation errors (63\%), whereas only 2\% of EvoSuite’s generated tests failed to run. 
Despite EvoSuite's relative success, our findings indicate that both tools are largely ineffective for this task, as they failed to generate any meaningful change-capturing tests for the large majority of the PRs (64\%).

Although in our evaluations, both generators could not achieve a high F2P ratio, and EvoSuite outperformed GPT-4o, we believe there are agentic code generation methods that hold a significant capacity.
Ultimately, our work highlights a critical gap in tooling and calls for the development of high-performance test generators tailored to the incremental nature of modern software development.

\end{abstract}

\begin{IEEEkeywords}
  software testing, automated unit test generation, large language models, pull requests, incremental test generation
\end{IEEEkeywords}

\section{Introduction}
For many years, the field of automated test generation has focused on creating test suites for entire code units, like classes \cite{fraser2011evosuite,fraser2017sbst,fraser2018sbst, shamshiri2015faults, holmes2020using}. 
The success of these methods is typically measured with metrics such as line/branch coverage over the entire file or mutation scores.
In practice, however, software systems often evolve through pull requests (PRs), which introduce small and specific changes to the code \cite{Zhang2023}.  
Industrial best practices emphasize the importance of unit tests in PRs to prevent regressions and maintain code quality~\cite{microsoft_engineering_playbook}.
When testing these changes, the goal is no longer about achieving broad coverage of the entire file. 
Instead, the focus shifts to creating tests that precisely test the new code and functionality \cite{ma2020commit}. 
This leads to the concept of fail-to-pass (F2P) tests \cite{jimenez2023swe}. 
An F2P test fails before the PR changes are applied (the base commit) and passes after the PR changes are applied (the head commit). 
This F2P criterion necessitates a new metric for evaluating PR-aware test generation.

While PR-aware test generation is a clear practical need, it remains underexplored in the literature. 
Prior work has approached the problem only indirectly, for instance through test generation from issue reports \cite{nashid2025issue2test, lou2024} or through regression and differential testing~\cite{zhou2025lwdiff, chen2016coverage,yoo2012regression,chen2023tevos, liu2023finerts, Liu2023, trautsch2023tevos}. 
% While PR-aware test generation is a clear practical need, this area has been overlooked by researchers.
% Existing research has explored this topic indirectly, such as generating tests from issue reports \cite{nashid2025issue2test, lou2024} or focusing on regression and differential testing \cite{zhou2025lwdiff, chen2016coverage,yoo2012regression,chen2023tevos, liu2023finerts, Liu2023, trautsch2023tevos}. 
However, these approaches do not directly address the unique challenge of PRs. 
To the best of our knowledge, no previous work has systematically focused on generating tests at the PR level with the explicit goal of creating fail-to-pass (F2P) test cases.

For the last decade, EvoSuite \cite{fraser2011evosuite} has been recognized as a leading tool for automated unit test generation. 
Numerous studies have confirmed its effectiveness, showing it achieves higher code coverage and finds bugs as well as, or better than, other tools like Randoop~\cite{pacheco2007randoop_oopsla_companion} and similar methods~\cite{panichella2021sbst_icse_companion}. 
This success has made EvoSuite the standard choice for generating Java tests in many research projects~\cite{fraser2011evosuite,fraser2017sbst,fraser2018sbst}.
However, EvoSuite is not without its drawbacks. 
A recent study by Gkikopouli et al.~\cite{gkikopouli2023empirical} pointed out that EvoSuite tends to create tests that confirm the software's current behavior, rather than tests that can uncover small, unexpected changes. 
These limitations are especially challenging when testing small code modifications, like those in a PR. 
This limitation highlights the difficulty of applying traditional whole-unit tools to the PR-aware context.

The recent emergence of Large Language Models (LLMs), with their significant capacities in code understanding and reasoning ~\cite{jimenez2023swe}, offers a promising new direction to address these challenges.
Recent research has explored the utilization of LLMs for unit test generation. 
For example, Jain et al. \cite{jain2025testforge} introduced TestForge, an agent-based generator that uses OpenHands~\cite{wang2025openhands}, which achieved promising results (44\% line coverage).
However, these promising results do not yet demonstrate a clear advantage over traditional methods.
A recent study by Abdullin et al.~\cite{abdullin2025test} compared LLMs to EvoSuite. 
Their findings indicate that traditional methods often still achieve better results in terms of both test coverage and fault detection (Section~\ref{sec:background-llm}).

Given the known limitations of established tools like EvoSuite for small-scale changes, coupled with the emerging potential of LLM, we are motivated to compare their performance on the task of PR-aware test generation.
We begin by collecting a dataset of PRs to serve as the evaluation benchmark for our study. 
While several public PR benchmarks exist, such as SWE-Bench \cite{jimenez2023swe}, Multi-SWE-Bench \cite{wang2024multiswebench}, and defects4j \cite{just2014defects4j}, they are incompatible with the strict technical requirements of EvoSuite. 
Specifically, EvoSuite exclusively supports the Java programming language and is fully functional only with version 8. 
This constraint, which we discuss in depth in Section~\ref{sec:data-collection}, necessitates the curation of a new dataset of PRs tailored for our study.

Our evaluation methodology is designed to measure how effectively each tool (Evosuite, GPT-4o) can produce (F2P) tests at the granularity of a single PR.
For EvoSuite, our process involves generating tests for an entire file that has been modified. 
Then, we execute the new test suite on the code version before the PR, checking for any F2P.
In parallel, we assess GPT-4o's capability at generating test cases for PRs. 
We provide the model with the PR's code difference (the diff), the whole source file, and any existing tests, and then ask it to generate test cases that specifically capture the introduced changes.
To further assess the usefulness of these tools, we categorize the PRs in our dataset by size. 
We then evaluate how well the tools perform for each size category.
This analysis helps us determine if the tools work better for certain PR sizes, which would mean they are more suitable for those specific situations.

\noindent
We summarize this paper's contributions as: 
\begin{itemize}
  \item Assessing the effectiveness of \textit{EvoSuite} for PR-aware test generation under a fail-to-pass objective;
  \item Evaluating single-prompt LLM prompting for PR-aware test generation, analyzing compilation success, and F2P yield; 
  \item Analyzing how the PR's size affects the effectiveness of each of the generators.
  \item Curating a new dataset to facilitate research in the field of test generation, focusing on the Java programming language.
\end{itemize}

\noindent The data and materials necessary to reproduce this study can be accessed at our replication package.\footnote{The data and materials necessary to reproduce this study can be accessed at \url{https://figshare.com/s/ddc3b1aacec45c30156b}}

\section{Background}
% In this section we talk about the related works in the test generation Domain. Initially, we look at the traditional automated test generation for Java, exemplified by search-based and feedback-directed methods. Afterwards, we talk about the recent LLM-powered approaches that generate or repair tests, including agentic variants.
\subsection{Traditional Automated Test Generation Methods}
Early approaches to unit test generation for Java focused on automated techniques like search-based and random test generation. 
Randoop \cite{pacheco2007randoop_oopsla_companion,pacheco2007feedback} is a prime example of a feedback-directed random testing tool that generates sequences of method calls, preserving those inputs that increase coverage. 
Such random techniques are aimed at quickly finding failing inputs or maximizing code coverage. 
On the other hand, EvoSuite employs search-based (genetic) algorithms to evolve test cases with the goal of covering as many branches as possible. 
EvoSuite has been shown to achieve high structural coverage and fault-detection effectiveness, often outperforming other tools in competition benchmarks \cite{fraser2011evosuite,fraser2017sbst,fraser2018sbst, panichella2021sbst_icse_companion}. 
In general, these classic tools succeed in producing extensive test suites with high coverage \cite{shamshiri2015faults}. 
However, a known drawback is that the resulting tests can be hard for developers to understand and maintain, due to unnatural sequences or obscure test oracles that require significant developer effort to debug \cite{shamshiri2018maintenance, palomba2016quality, pecorelli2024granular}.

\subsection{LLM-Powered Test Generation}
\label{sec:background-llm}
Recent work explores using LLMs to generate unit tests, given the impressive code generation abilities of models like GPT-4. 
A straightforward single-prompt approach is to supply the code under test to an LLM and request unit tests. While this often yields more human-like and readable test code compared to search-based tools, in practice, it has significant reliability issues. 
Studies have found that a large portion of tests generated by off-the-shelf LLMs are malformed or non-compiling, often due to hallucinated dependencies or incorrect assertions \cite{yang2024empirical,sapozhnikov2024testspark, yuan2023no}. 
Hallucinated test setups and outputs frequently cause these single-prompt generated tests to fail when run, severely limiting their usefulness without manual fixes \cite{yang2024empirical}. 
To address such issues, researchers have developed enhanced LLM-driven testing tools that incorporate feedback and iteration.

TestSpark (2024)~\cite{sapozhnikov2024testspark} is an IntelliJ IDEA plugin that integrates both search-based generation and LLM-based generation for Java. 
Notably, TestSpark introduces a feedback loop between the IDE and the LLM to ensure any tests the model produces are actually compile-safe.
In this approach, after the LLM proposes test code for a Java class or even a specific line, the tool checks compilation and can prompt the LLM to correct errors, ensuring all final generated tests compile successfully. 

TestForge~\cite{jain2025testforge} offers another iterative, feedback-driven refinement process, using OpenHands~\cite{wang2025openhands} to generate test cases for Python code. 
Instead of relying on a single-prompt, it frames generation as a multi-step loop: the model creates initial tests, executes them, analyzes failures and coverage gaps, and then revises the suite until the desired quality is achieved. 
This enables self-correction across iterations, producing test suites that are both high-coverage ($\approx 44\%$ line coverage) and high-quality, often more natural and understandable—closer to human-written tests—than those from search-based generators. 

In summary, the landscape has shifted from single-prompt test generators (e.g., random or evolutionary methods) to intelligent, hybrid approaches leveraging LLMs. 
Early methods like EvoSuite and Randoop delivered strong coverage and fault-finding ability \cite{fraser2011evosuite,fraser2017sbst,fraser2018sbst,shamshiri2015faults}, but at the cost of maintainability and ease of use \cite{shamshiri2018maintenance}. 
The latest LLM-driven techniques – from single-prompts to advanced systems like TestSpark and TestForge – focus on generating more developer-friendly tests. 

\subsection{Test Generation for PRs and Commits}
Although numerous studies have explored regression testing at the commit or patch level \cite{ma2020commit, xiang2024critical, liu2025can, IdentifyandUpdate}, to the best of our knowledge, none have directly investigated LLM-based unit test generation at the granularity of PR.
Xiang et al.~\cite{xiang2024critical} introduced WAFLGo, a directed greybox fuzzer designed to detect vulnerabilities introduced by new commits. 
WAFLGo guides input generation toward recently modified and data-dependent code regions using a critical-code-aware distance metric, allowing it to thoroughly explore the affected areas and reveal commit-induced regressions.
Building on this idea, Liu et al.~\cite{liu2025can} presented Cleverest, a feedback-directed, LLM-based regression test generator that extends the WAFLGo paradigm from fuzzing to natural-language-guided input synthesis. 
Given a commit message or code diff, Cleverest iteratively prompts an LLM to craft test inputs that differentiate the behavior of pre- and post-commit versions. 
In their conclusion, the authors emphasize that LLMs can automatically generate—or at least bootstrap the manual creation of—regression test cases for code commits, specifically in scenarios where the target programs accept highly structured, human-readable inputs (e.g., JavaScript or XML) as their test cases. 
While these tools successfully expose behavioral differences at the input level, they do not produce compilable, source-integrated unit tests. 

\subsection{F2P Criterion and SWE-Bench}

Recent benchmarks such as SWE-Bench~\cite{jimenez2023swe} have popularized the F2P criterion, which measures whether a test fails before a PR (base commit) and passes after the PR (head commit). 
In SWE-Bench, this principle defines success: the patched system is considered correct only if the tests that were failing in the faulty version become passing after the patch is applied. 
This gives a direct, behavior-focused measure of correctness rather than using indirect metrics like coverage.
F2P is particularly important because it captures the true intent of software evolution. 
Traditional measures like line or branch coverage quantify how much code is executed, but not whether the generated artifacts validate a specific change. 
The F2P property instead reflects semantic alignment—if a test transitions from fail to pass, it necessarily exercises the behavior introduced by the change. 
For this reason, our study adopts the F2P criterion, as a direct and reliable measure of whether generated tests genuinely detect and confirm PR modifications.

\section{Methodology}
\label{sec:methodology}
We arrange the content of the study around three research questions. 

\global\def\RQOne{RQ1: Is \textit{EvoSuite} effective at generating F2P tests cases for PRs?}
\textbf{\RQOne{}}

We measure the fraction of PRs for which EvoSuite produces at least one test that fails on the base commit and passes on the head. In this research question, although the quality and coverage of the generated test cases are important, we take one step back and focus on a simpler but still meaningful metric. Instead of evaluating how comprehensive the test cases are, we aim to determine whether there is at least one generated test case that directly addresses the newly introduced changes.

\global\def\RQTwo{RQ2: Is \textit{GPT-4o} effective at generating F2P tests cases for PRs?}
\textbf{\RQTwo{}}

For this research question, we measure the effectiveness based on two factors. 
First, what is the ratio of the test cases that are compilable? 
And second, do they contain any F2P tests or not (Section~\ref{sec:evaluation})? 
The details about the prompts and the provided context are discussed in Section~\ref{sec:promptEngineering}.

\global\def\RQThree{RQ3: How does the size of a PR, affect the effectiveness of EvoSuite and GPT-4o in generating tests?}
\textbf{\RQThree{}}

PRs are not all the same. 
They range from simple, single-file typo fixes to complex features that change many files. 
This difference in size and complexity might affect our two generators differently.
While the number of files modified in a PR does not fully determine its size and complexity, it is an indicator of the overall scale of the change.
To understand this, we connect the F2P success rate with the number of files in the PR to understand when the generators work best.

\begin{figure}[t]
  \includegraphics[width=\columnwidth]{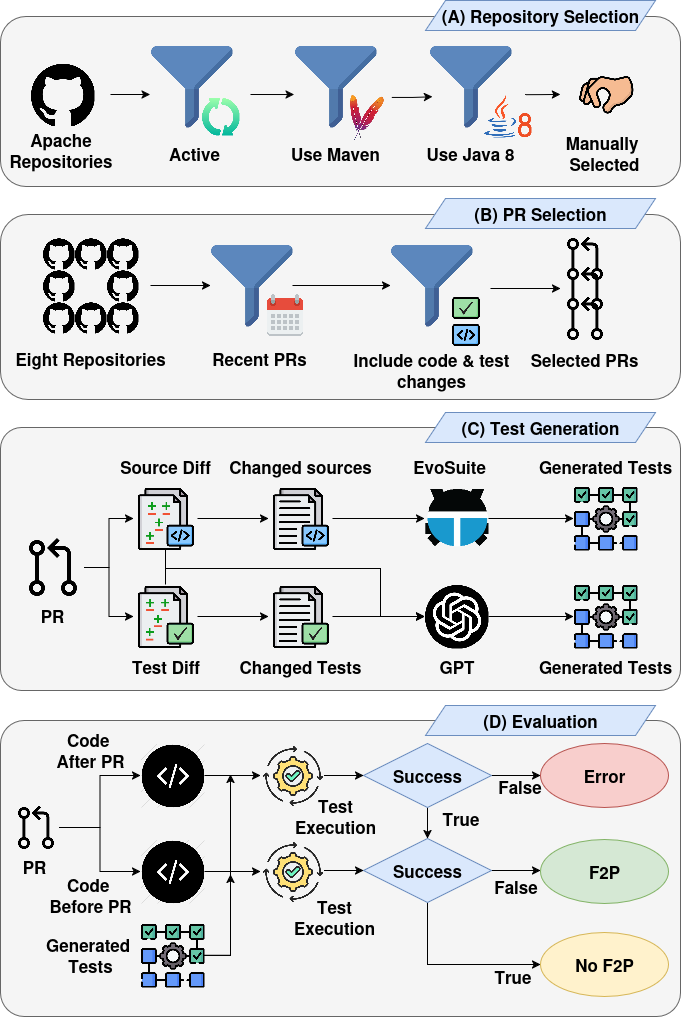}
  \caption{Demonstration of the overall study process}
  \label{fig:methodology}
\end{figure}

To address these research questions, an overview of our methodology is presented in Figure~\ref{fig:methodology}.

\subsection{Dataset Collection}
\label{sec:data-collection}

Despite the growing trend in the literature towards code generation for PRs, which has led to the creation of several datasets containing PRs and their corresponding issues, such as SWE-Bench~\cite{jimenez2023swe}, we found that existing datasets were not suitable for the purpose of our study.
The first challenge we encountered is that the majority of these datasets, including the original SWE-Bench, focus on the Python programming language. 
This is a significant limitation for our work, as we aim to evaluate EvoSuite, a tool that exclusively supports Java.
Although a few datasets do include Java, such as Multi-SWE-Bench~\cite{wang2024multiswebench}, SWE-Bench-Multilingual~\cite{yang2025swesmith}, and SWE-bench-java~\cite{zan2024swe}, they present other problems. 
The size of these datasets is substantially smaller compared to the original SWE-Bench, and they include few to no tasks designed for Java 8.

The requirement for Java 8 is critical because it is the version that EvoSuite most reliably supports. 
While support for Java 9 has been mentioned, there are not many available repositories developed with that version. 
We also investigated claims of Java 11 support, but our own experiments consistently crashed with errors indicating that the API version is incompatible.
Therefore, to properly evaluate EvoSuite's ability to generate tests, we needed to collect a new dataset consisting of PRs from projects specifically using Java 8, creating a suitable environment where EvoSuite is not at a disadvantage. 

\paragraphTitle{Repository Selection}
\label{sec:repoSelection}
We collect the initial dataset for our evaluation from the well-established open source libraries from the Apache Software Foundation \cite{apache2025}. 
We filter the repositories to have at least 10 maintainers, which can be an indicator of proper PR culture.
The resulting selection includes repositories with a significantly higher number of contributors, specifically a minimum of 49.
Then, we sort the repositories based on the number of PRs between January 1, 2019, and 2024, to select the ones that have been active recently. 
In this context, we define “active” as repositories that have shown consistent maintenance and community engagement during this period, such as ongoing PR activity, recent merges. 
This particular time frame is chosen to ensure compatibility of Java and the major JUnit versions.
For simplicity in our evaluation, we select only repositories that use Maven \cite{maven} as the build system. 
Then, we filter the list to include only projects that use Java as their main programming language, making sure that Java files make up over 95\% of the code.

After this, we manually review the remaining repositories to further narrow down our selection to those specifically using Java 8.
We also select repositories with a single module whose test folder is in the standard location under \texttt{<project-root>/src/main/java}. 
This setup allows us to locate test cases easily and link them to their corresponding source files, which we send to \llm{} as part of the context.
We end up with eight repositories listed in the Table~\ref{table:projects}.
An abstract overview of the process is demonstrated in the Figure~\ref{fig:methodology}-A.

\paragraphTitle{PR Selection}
To build a dataset of the PRs, we curated a set of PRs through a series of steps.
First, we collect all PRs opened between January 1, 2019, and the end of 2024. 
From this initial pool, we select only the PRs that were successfully merged into the main branch. 
We then filter this set further, keeping only PRs that added or modified at least one Java file. 
This step was important to exclude changes that were not relevant to our study, such as dependency bumps or files that were only renamed or deleted.
After these initial filters, we perform a manual review to refine the dataset. 
We began by investigating where EvoSuite fails to generate tests. 
Because EvoSuite is known for its low error rate~\cite{abdullin2025test}, we hypothesize that these failures might stem from existing errors within the PRs themselves, rather than from an issue with EvoSuite.
We further investigate all cases where our pipeline fails to run the tests or could not get the tests from GPT-4o.

% Finally, we remove 15 PRs that only made changes to Java interfaces, abstract classes, as these cannot be tested directly, and our generators (EvoSuite and GPT-4o) were unable to generate tests for them. 
% We also remove four PRs because they were only touching comments in Java files. 
% We also exclude seven PRs because they contain files that were too large for GPT-4o's input limit. 
% For the sake of simplicity, we chose to focus on the other PRs. 
% Additionally, we remove three PRs where we can not link a source code file to its corresponding test file, ensuring the dataset remained consistent.
To manage the execution time and technical complexity of our analysis, we limit the number of PRs to a maximum of 50 per project. 
When a project has more than 50 eligible PRs, we prioritize the most recent ones. 
While not all projects reach this limit, this cap helps create a more balanced dataset.
These filtering steps result in a final set of 353 PRs. 
A detailed distribution of these PRs across the different projects is presented in Table \ref{table:projects}.
A general overview of our PR selection is demonstrated in the Figure~\ref{fig:methodology}-B.

\begin{table}[t]
    \centering
    \resizebox{\columnwidth}{!}{%
        \begin{tabular}{l l c c c}
            \toprule
            \textbf{Project Name} & \textbf{Stars} & \textbf{Contributors}  & \textbf{Total PRs} & \textbf{Fetched PRs} \\
            \midrule
            commons-lang & 1.7K & 235 & 1362 & 48 \\
            commons-io & 1k & 121 & 754 & 50 \\
            commons-collections & 708 & 81 & 609 & 49 \\
            commons-codec & 469 & 50 & 388 & 33 \\
            commons-imaging & 463 & 45 & 508  & 50 \\
            commons-csv & 393 & 49 & 562 & 32 \\
            commons-compress & 374 & 98 & 661  & 50 \\
            commons-text & 363 & 63 & 676 & 41 \\
            \bottomrule
        \end{tabular}%
    }
    \caption{List of the projects selected for experiment, including the corresponding number of fetched PRs}
    \label{table:projects}
\end{table}

\subsection{EvoSuite Test Generation}
\label{sec:evosuite}
EvoSuite is widely recognized in many studies as a state-of-the-art, search-based test generator. 
Although the official EvoSuite website still provides documentation for a regression-testing feature that was available in earlier versions \cite{evosuite_web}, our findings indicate this feature is no longer supported in recent releases. 
Specifically, when attempting to use the documented \texttt{-regressionSuite} command, the tool consistently fails and returns the following error: "ERROR EvoSuite - Parsing failed. Reason: Unrecognized option: -regressionSuite". 
We first observed that the command for regression testing was unrecognized and that no alternative instructions were available. 
This finding was supported by a discussion in an official repository issue, which confirmed that this feature had been removed~\cite{EvoSuiteIssue353}. 
Based on these two pieces of evidence, we concluded that the regression testing functionality has been discontinued in the newer versions of EvoSuite.

To overcome the absence of a dedicated regression testing mode, we use EvoSuite to generate tests for each modified class after the PR (the head of a PR). 
We then identify a test as relevant if it fails before the PR (the base of the PR) but passes after the PR. 
This method allows us to isolate the specific tests that validate the newly introduced changes, which we refer to as fail-to-pass tests. 
Since EvoSuite's primary goal is to maximize coverage, in theory, it is expected to also generate tests that covers the lines and branches newly introduced by the PR.
EvoSuite also has another limitation that it can generate tests for one class at a time. 
Because a typical pull request often involves changes to multiple classes, our process involves running EvoSuite separately for each affected class.

\paragraphTitle{Test Generation}
For our study, we configured EvoSuite with specific settings to ensure optimal results. 
We set the search budget to 120 seconds per class, which is double the default of 60 seconds. 
We found that this extended time allows for better test generation, especially for larger files. 
We also experimented with longer budgets, such as 180 seconds, but observed minimal to no improvement, making 120 seconds the most efficient choice.
Also, we let EvoSuite use the DynaMOSA test-generation algorithm, which is the default option.
Our methodology for generating and executing tests with EvoSuite follows a clear, multi-step process.
First, we build the project using its original $package$ target in Maven. 
This step produces the necessary JAR file that EvoSuite needs to analyze the code.
Next, we run EvoSuite using the generated JAR file, specifying which class to generate tests for. 

For each target class, EvoSuite produces two key classes within a nested folder structure that mirrors the class's package.
One is a Test Class, which contains the actual test cases.
The other one is a Scaffolding Class, which sets up the required runtime environment for the tests. 
The generated class inherits the scaffolding class, which will automatically initialize the runtime environment for the test during execution. 
This scaffolding also creates a secure sandbox environment to prevent the tests from making unintended changes to the system.
All test classes generated by EvoSuite have an \texttt{\_ESTest} or \texttt{\_scaffolding} suffix added to their filenames, which does not produce any conflict with already existing test classes in the project. 
We then move these generated files into the project's main test folder, placing them alongside the original test files.
Finally, to run only the tests generated by EvoSuite, we use Maven with a wildcard that targets the unique suffix. 
By executing only tests matching \texttt{*\_ESTest}, we can precisely measure the behavior of the EvoSuite-generated tests while ignoring all other tests in the repository.

\subsection{Single-Prompt LLM Test Generation}
To evaluate the capacity of LLMs, we chose OpenAI's GPT-4o~\cite{openai_gpt4o_system_card_2024} , as it is one of the widely used LLMs. 
Our approach involves generating tests using a single-prompt that provides GPT-4o with three inputs for each modified file: the pull-request diff, the complete source file, and any pre-existing tests associated with that source file. 
We intentionally exclude the new or modified tests from the PR itself to avoid showing the model the expected answer.
During our experiment phase, we observed that by providing already existing tests and explicitly instructing the model in the prompt not to create tests that overlap with the provided test cases, we can better guide it to focus on generating tests for the new changes introduced in the pull request.
For this experiment, we set the temperature to 0.1 to obtain more deterministic results and limited the maximum output length to 64K tokens.

As described in Section~\ref{sec:data-collection}, we select repositories that allow us to reliably map each source file to its respective test file.
To stay within the model's context limit, we process each pull request on a file-by-file basis, generating tests for each file independently. 
Still, we need to exclude seven PRs because they were touching larger files that do not fit in the context window. 
The generation process only targets Java files that are added or modified. 
We skip deleted or renamed files, as they do not introduce new logic that requires new test cases.
Initially, we experimented with making five attempts to generate tests for each file. 
However, we observed that having more attempts rarely improved the outcome, which was mainly failing due to compile error. 
If the model is hallucinating (e.g., assuming a specific class or method exists when it does not), it would typically repeat the same error in subsequent attempts. 
Since multiple tries did not significantly change the results, we decided to limit the generation to a single attempt. 
This decision allowed us to limit the costs of the study and, in turn, evaluate a larger dataset.

\subsubsection{Prompt Engineering}
\label{sec:promptEngineering}
We aim to optimize the model's output by refining the prompt we used for generation. 
To accomplish this, we experimented with several prompt modifications, evaluating their impact on a sample of 20 PRs.
Initially, we considered providing the model with a one-shot example, which includes a sample file and the desired changes (a diff). 
We observed that using an example with a small file and a small diff did not produce significantly better results. 
This is because most of the PRs introduce very small changes to large files. 
However, using a large file as an example would lead to high computational costs and potential problems with the model's context window, so we did not pursue that approach.
We then shifted our focus to the model's reasoning process. 
We tried asking the model to first describe the necessary changes in a "chain of thought"~\cite{wei2022chainofthought} before generating the tests. 
This, however, did not meaningfully improve the outcome. 
The primary obstacle remained the compilation of the generated tests. 
We also attempted to have the model validate its own output and make sure it is compilable, but this was unsuccessful; if the model generated a test with a compilation error, it would continue to make the same mistake repeatedly.

Also, we encountered a recurring issue where the model-generated code would fail with a runtime error. 
This problem occurred more frequently in software repositories such as \texttt{imaging}. 
These runtime errors occurred primarily because such libraries depend on external files for input or output, and the generated tests attempted to access files that were not present.
To solve this problem, we experimented with several strategies.
First, we tried providing the model with a list of available files from the project's resources folder, instructing it to only use those files. 
This approach did not lead to significant improvement. 
The model either continued to use files that were not on the list or used an available file, but the provided files did not function as the test expected. 
From this, we speculate that the model needs to understand the content of the files, not just their names, to choose a suitable input.

Our next approach was to instruct the model to not use any external files at all. 
However, this led to a higher number of compilation errors and a noticeable decrease in the number of generated tests that executed successfully. 
The model then tried to create inputs for functions by inventing new classes that were not part of the project.
Finally, we directed the model to use mocking (specifically with Mockito) to create test inputs instead of using real files. 
However, we saw the same pattern of compile errors and a reduction in the number of successfully executed tests compared to the previous configuration
Overall, we could not find a solution to this problem. 

\paragraphTitle{Helpful Methods}
Through this process, we found a few key modifications that helped. 
First, we specifically requested in the prompt that the model provide all the imports required to run the test class. 
This simple instruction helped improve the issue with compilation errors and solved a small portion of the problem.
Second, we instructed the model to generate test classes within the same package as the source code, simply adding an \texttt{\_LLMTest} suffix to the new class name. 
Although this method showed a significant improvement in the results, issues still occurred frequently during the evaluation phase. 
To solve this problem, we created scripts to automatically update all package and class names, which helped prevent these potential errors. 

Finally, we analyzed how the model presented its output. 
We observed that the model was much better at providing only the code when asked, compared to when we requested a structured output with text explanations. 
This change made it much easier for us to automatically extract the generated tests. 
Using this method, we only encountered seven cases where our automated pipeline could not extract the tests. 
Although we could have regenerated these, we chose to count them as errors to ensure a fair evaluation. 
The final version of the prompt we developed through these experiments is shown in Figure~\ref{fig:prompt_template}.

\begin{figure}[h!]
    \centering
    \begin{tcolorbox}[breakable,
        colback=gray!5,
        colframe=black!20,
        fontupper=\ttfamily\small,
        arc=1mm,
        boxrule=0.3pt,
        left=6pt,right=6pt,top=6pt,bottom=6pt]
As a Java test engineer, your role is to update or create new unit test cases based on the changes in a pull request. Utilize the old version of the code and the diff file to understand and cover the changes in your tests.

Input 1 – Old Version of the Code: ...
Input 2 – Diff File: ...
Input 3 – Existing Test Case Class: ...

Determine the name for the new test class by appending 2 to the existing test class name. For instance, MyClassTest becomes MyClassTest2.\\

Instructions:
\begin{enumerate}
  \item Begin your output with the package declaration. You can infer the package name from the given file paths.
  \item Include necessary Java imports at the beginning of your response.
  \item Carefully analyze the diff file to identify what specific changes have been made to the code and what new functionalities or modifications need to be tested.
  \item Generate tests that target these changes specifically. Ensure these new tests do not duplicate any functionality of tests found in the existing test case class. Refer to the existing test class to understand what tests are already in place and identify gaps in coverage.
  \item Include comments and annotations in your code to increase readability.
  \item Do not include any natural language descriptions in your response.
\end{enumerate}
    \end{tcolorbox}
    \caption{Prompt Template}
    \label{fig:prompt_template}
\end{figure}

\subsection{Experiment Setup}

For each PR, we begin by integrating all the generated test files. 
Since a single PR can involve multiple classes, this may result in several test files. 
We combine these into the main test folder of the project. 
To ensure proper organization, we designed the generation process so that new test classes are placed within the same package as the code they are testing. 

This merges them into the correct package folder alongside the project's original tests. 
To prevent naming conflicts with existing test classes, we add a specific suffix, either \texttt{\_ESTest} or \texttt{\_LLMTest}, to each new test class name.
Once the test suites are integrated, we evaluate their quality in a two-step process. 
First, we verify that the generated tests are functional. 
To do this, we require that every test case compile and run successfully on the head commit of the PR (the version of the code after the changes have been applied).
Next, we check if the test suite effectively captures the changes made in the PR. 
We do this by looking for F2P cases. 
A test is considered F2P if it passes on the head commit but fails on the base commit (the version of the code before the PR changes). 
This indicates the test successfully covers the new behavior or fix introduced by the PR.
Figure~\ref{fig:methodology} (C and D) demonstrate a summary of this process.

During our study, we encountered a few specific situations that required special handling.
One important case involves PRs that introduce new methods or change existing method signatures. 
For these PRs, the generated tests pass on the head commit as expected, but then fail to compile on the base commit, since the new or modified code does not exist yet. 
We consider these tests to be effective F2P cases because the compilation error directly captures the structural changes introduced by the PR. 

Another common scenario arose with test suites generated by EvoSuite, which were often very large and sometimes contained a few tests that failed on the head commit due to assertion errors or runtime exceptions. 
Instead of discarding the entire suite, which was a frequent issue, we adopted a more practical approach. 
We simply ignored the individual failing tests and proceeded with the valid ones, as long as at least one passing test case remained in the suite.
However, a limitation of this approach is that in case of compile error we cannot pinpoint which specific test within a suite causes a compilation error so that we can remove it. 
As a result, if any compilation error occurs, the entire suite is marked as "Error."

For our experiments, we execute tests using Surefire \cite{apache_maven_surefire_plugin_docs}, which is the standard test-execution plugin for Maven. 
To execute and evaluate the tests, we place the generated test cases in the project’s test directory, under the package folders where they belong. 
To prevent any conflicts with existing files or classes, we add a specific postfix to the name of each generated test case, such as \texttt{\_ESTest} or \texttt{\_LLMTest}. This guarantees that every new test file has a unique name.
This unique naming convention also helps us optimize the test execution time. Instead of running a project's entire test suite, we configure Surefire to run only our generated test cases. 
We achieve this by instructing it to execute only those tests whose class names end with our specific postfix (for example, \texttt{*\_ESTest}). This approach significantly reduces the overall runtime.

\subsection{Evaluation}
\label{sec:evaluation}
To measure the effectiveness of the test generators, we classify each generated test into one of the following three categories. 
The performance of the generators is then evaluated based on the percentage of tests that fall into each group.
Figure~\ref{fig:methodology}-D provides an overview of this classification.

\begin{itemize}
\item \textbf{Error}: This category is for tests that are invalid from the start. 
It includes tests with compilation errors, cases where the tool crashes and fails to produce a test file, or situations where all generated tests fail immediately after being created.
\item \textbf{No F2P}: This group contains tests that compile and run correctly but are not useful. 
They successfully run both on the head and the base of the PR.
Therefore, these tests do not check any of the new changes in the Pull Request (PR) and are unrelated.
\item \textbf{F2P}: These are the useful tests that successfully capture an aspect of the PR's changes. 
A test is considered an F2P if it fails on the base commit in one of two ways: by causing a compilation error (for example, when a test expects a new function that does not exist in the older code) or by an assertion failure (when the test expects different program behavior due to the new logic).
\end{itemize}

\noindent
To answer RQ1 and RQ2, we take each PR and assign its generated tests to one of the three categories described above. 
We then report the percentage of tests in each category for every repository and as an overall total.
Initially, we calculated the F2P ratio across the entire dataset, which included all generated outputs. 
However, we observed that GPT-4o produced a notably high number of errors. To understand the performance without the influence of these errors, we performed a second round of calculations. 
In this subsequent analysis, we recalculated the F2P percentage after excluding the error cases from both generators.
For a deeper analysis to answer RQ3, we also examine how the number of files in a PR affects the generation of useful (F2P) tests. 
To do this, we group PRs into four bins based on the number of modified Java files: 1 file, 2–3 files, 4–5 files, and more than 5 files. 
The number of PRs in each of the bins is represented in Table~\ref{table:bins}.
We then measure the success rate of the test generators, which we define as the percentage of PRs for which at least one useful F2P test was produced.

\begin{table}[htbp]
    \caption{Number of PRs broken down to the number of files they change}
    \label{table:bins}
    \centering
    \resizebox{0.4\columnwidth}{!}{%
        \begin{tabular}{c c}
            \toprule
            \textbf{File Count} & \textbf{PR Count}\\
            \midrule
            1 & 200 \\
            2-3 & 73 \\
            4-5 & 27 \\
            5+ & 53 \\
            \bottomrule
        \end{tabular}%
    }
\end{table}

\section{Results}
\begin{figure}[t]
  \centering
  \includegraphics[width=\columnwidth]{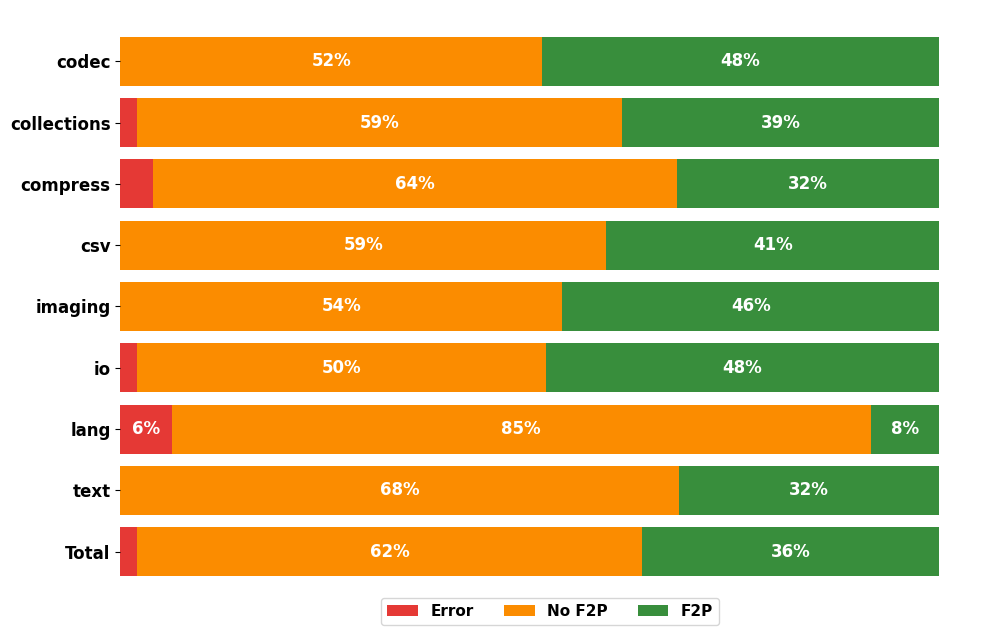}
  \caption{Evaluation Results for EvoSuite}
  \label{fig:results-evosuit}
\end{figure}
\label{sec:results}

This section presents the results of our evaluation and the answers to our research questions.
\subsection{\RQOne{}}
The evaluation results for EvoSuite are summarized in Figure~\ref{fig:results-evosuit}.
Overall, our findings show that EvoSuite is highly reliable, as only 2\% of its generated tests resulted in compile errors. 
However, for a majority of the PRs, specifically 62\%, the generated test suites were ineffective because they did not contain any F2P test cases.
For the remaining 36\% of PRs, EvoSuite successfully generated a test suite that included at least one F2P test case.
At the end, if we overlook the tiny percentage of the error, EvoSuite could generate at least one F2P in 37\% of the cases.
A notably high ineffectiveness has been observed in the \texttt{commons-lang} repository. 
Specifically, EvoSuite encountered a "no F2P" rate of 91\% for this project. 
Through manual evaluation, we speculate this increase occurred because the majority of files in this repository are very large, often consisting of thousands of lines of code. This puts EvoSuite at a disadvantage, since it needs to generate tests for the entire file.

\begin{tcolorbox}[colback=gray!10!white, colframe=gray!75!black, title=EvoSuite Evaluation]
Although EvoSuite achieved a high compilability rate of 98\%, 62\% of the generated tests were not containing F2P for the PRs, and only 36\% contained at least one F2P test case (37\% if excluding errors).
\end{tcolorbox}

\begin{figure}[t]
  \centering
  \includegraphics[width=\columnwidth]{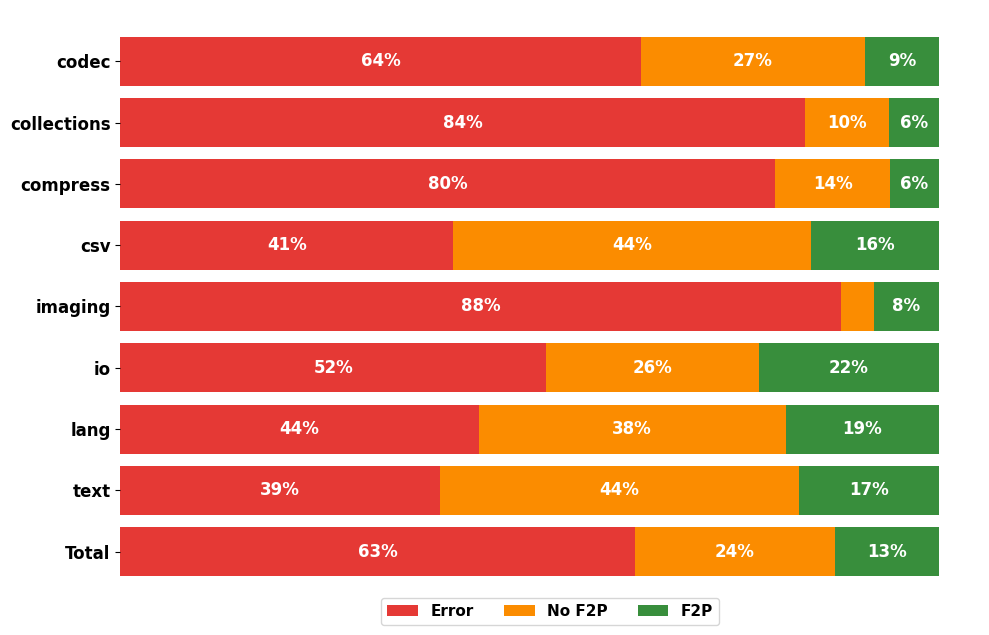}
  \caption{Evaluation Results for GPT-4o}
  \label{fig:results-llm}
\end{figure}

\subsection{\RQTwo{}}
The evaluation results for GPT-4o are summarized in Figure~\ref{fig:results-llm}.
Our evaluation of GPT-4o shows that 63\% of the generated test suites contained Errors. 
For the remaining attempts, GPT-4o successfully created an F2P test in 13\% of cases and produced Ineffective test cases in 24\% of cases.
If we disregard the instances with errors and only consider the successful and ineffective generations, the rate of F2P creation rises to 34\%. 
This performance is slightly lower than that observed with EvoSuite (37\%).
We observed that the high rate of errors was not consistent across all projects. 
The compile errors were especially frequent for certain projects, such as \texttt{commons-imaging}. 
This project's nature involves heavy file manipulation, which led GPT-4o to generate a higher number of compile errors (Section~\ref{sec:promptEngineering}). 
The primary cause was that the generated test cases frequently failed when attempting to access files that did not exist.

\begin{tcolorbox}[colback=gray!10!white, colframe=gray!75!black, title=\llm{} Evaluation]
GPT-4o's test generations resulted in an Error 63\% of the time, which raises a concern about its stability. It produced a successful F2P in 13\% of all cases. When errors are excluded from the calculation, the F2P success rate increases to 34\%.
\end{tcolorbox}

\subsection{\RQThree{}}
\begin{figure}[t]
  \centering
  \includegraphics[width=\columnwidth]{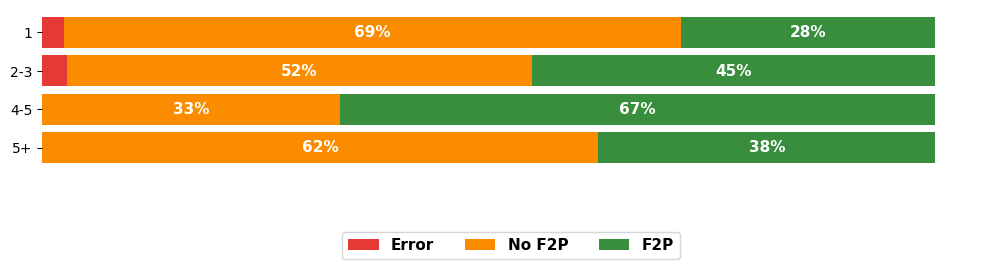}
  \caption{EvoSuite's PR performance broken down by the number of files}
  \label{fig:results-evosuit-bins}
\end{figure}

\begin{figure}[t]
  \centering
  \includegraphics[width=\columnwidth]{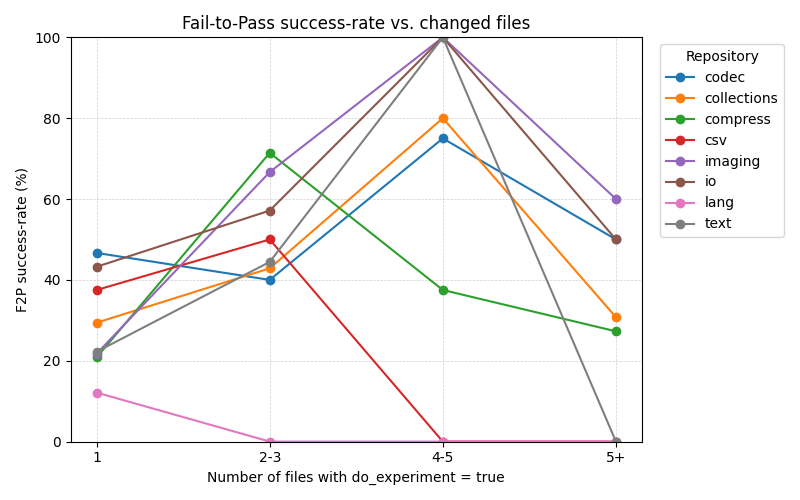}
  \caption{EvoSuite's success rate in different projects broken down by the number of files}
  \label{fig:results-evosuit-bins-project}
\end{figure}

The results for EvoSuite, broken down by the number of files, are presented in Figure~\ref{fig:results-evosuit-bins} and \ref{fig:results-evosuit-bins-project}.
Overall, we see a clear trend for EvoSuite: as the number of files in a PR increases, the F2P ratio tends to rise. 
A larger PR provides more opportunities for EvoSuite to generate tests, which in turn increases the probability of achieving at least one F2P. However, this positive trend does not continue indefinitely. 
After reaching a certain number of files (4-5), the trend reverses, and the F2P ratio begins to fall.

\begin{figure}[t]
  \centering
  \includegraphics[width=\columnwidth]{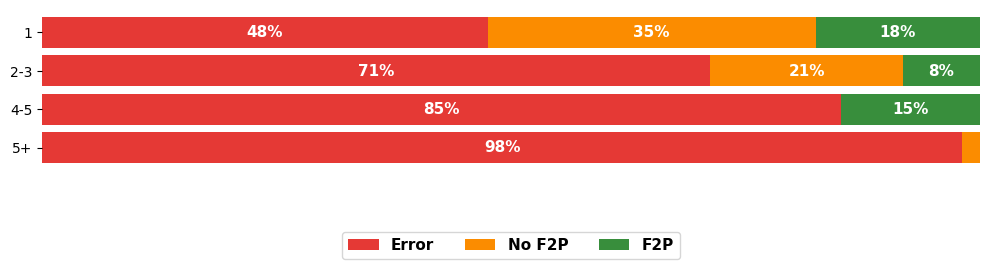}
  \caption{GPT-4o's success rate in different projects broken down by the number of files}
  \label{fig:results-llm-bins}
\end{figure}

\begin{figure}[t]
  \centering
  \includegraphics[width=\columnwidth]{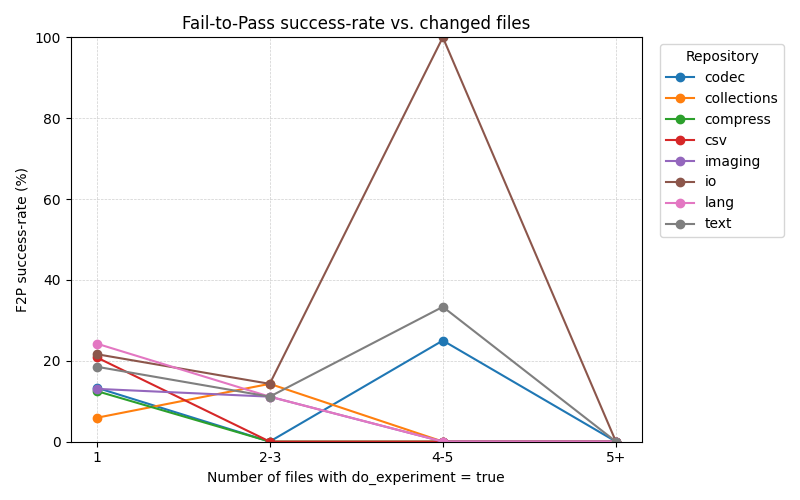}
  \caption{GPT-4o's PR performance broken down by the number of files}
  \label{fig:results-llm-bins-project}
\end{figure}

In contrast, GPT-4o shows a different pattern, as summarized in Figure~\ref{fig:results-llm-bins} and \ref{fig:results-llm-bins-project}.
For GPT-4o, an increase in the number of files corresponds to a growth in compile errors. 
Although the F2P ratio fluctuates with the size of the PR, the steady rise in the compile-error rate points to a drop in overall performance. 
This occurs because generating a larger volume of tests creates more chances for GPT-4o to introduce mistakes, which negatively impacts its effectiveness.

\begin{tcolorbox}[colback=gray!10!white, colframe=gray!75!black, title=File Count Effect]
As the number of files increases, EvoSuite gains performance up to a certain point; however, GPT-4o will suffer from more compile errors and experience a decrease in performance.
\end{tcolorbox}

\section{Discussion}
\subsection{Implications of the results}
\paragraphTitle{EvoSuite}
Our evaluation shows that while EvoSuite often produces test suites that compile, many are ineffective because they do not target the specific changes in a Pull Request (PR). 
In this study, we used a very generous rule to label a test suite as F2P.
We marked a suite as F2P if at least one test case touched any part of the PR's changes. 
This standard is much lower than industry standards to have strong branch and line coverage to fully test the new changes \cite{rojas2015combining}.

Even with this generous metric, we found that 62\% of the generated test suites failed to capture the PR changes. 
This indicates that EvoSuite is not well-suited for PR-aware test generation. 
We believe the main reason for this is EvoSuite's design. 
It is built to generate tests for the entire project, not to focus on a small, specific parts of the code where changes happen. 
This problem is worse in large Open-Source Software (OSS) projects. 
In these projects, files can be huge, while the PR changes only affect a tiny portion of the code. 
In our manual evaluation, we observed that as the file size grows, the likelihood of EvoSuite generating ineffective tests (No F2P) increases. 
While this relationship could be a separate study, we conclude that, EvoSuite is not well-suited to our task and settings.

\paragraphTitle{GPT-4o} 
In contrast, GPT-4o struggled to produce usable test cases that do not result in an error. 
Many of its generated tests could not be compiled or run without runtime errors. 
In total, 63\% of the PRs are labeled as 'Error' status.
This high error rate directly limited the model's ability to provide F2P tests. 
Even in many of the generated test suites that were not marked as 'Error', most of the test cases failed with a runtime error, leaving only a small number of executable tests. 
This skewed the results, increasing the 'No F2P' count. 
A closer review showed many of these 'No F2P' cases should have been labeled as 'Error'. 
However, we kept our evaluation method consistent for all tools. 
Changing it would have been unfair to EvoSuite, as we would not have counted its partially working tests.

We experimented with many prompt engineering techniques (Section~\ref{sec:promptEngineering}). 
We observed that, GPT-4o did not always follow our instructions to focus on the PR changes, resulting in 'No F2P' tests. 
This might be because of adding entire class and its corresponding test class in the prompt. 
This large block of code may have "dominated" the prompt. 
However, not sending the class body led to even more compile errors. 
Hence, we found that including the class was the better choice between these two options.
After many trials across different repositories, we are convinced that a single-prompt approach is not reliable enough for PR-aware test generation. 
Even a well-designed "best" prompt often failed, as repository-specific issues would arise. 
We observed that a prompt that worked well in one repository did not necessarily work for another.

\paragraphTitle{Number of modified files}
A manual evaluation of our dataset revealed that Pull Requests (PRs) affecting more than five files often involve a large number of files (e.g., 20). 
Yet either they introduce only a small change within each file or the PR contained a big feature implementation. 
For instance, some of these PRs are simple refactoring tasks, such as renaming a method that is referenced in many locations. 
This action leads to a high number of changed files, even though the underlying modification is minor.
Based on this observation, we believe that automatic test generation may offer less value for PRs in this "5+ files" category. 
There are two likely reasons for this. 
First, the PR might represent a major feature implementation with too many changes, which should ideally be broken down into smaller chunks for test generation.
Second, the changes are tiny and scattered (like the renaming example), making them less impactful. 
Furthermore, our results indicate that the test generation tools do not perform well in these specific scenarios.

\subsection{Future Directions}

As we discussed in the previous section, we see that a single prompt has a low chance of being able to capture the needs for PR-aware test generation. 
However, there are studies \cite{jain2025testforge, Konstantinou2025YATE, abdullin2025test} reporting large improvements in mitigating compilation errors in the tests generated by LLMs by using a feedback loop that lets agents assess and revise the generated tests. 
Konstantinou et al.~\cite{Konstantinou2025YATE} introduced \textsc{YATE}, a feedback-driven repair framework that recompiles and fixes LLM-generated tests using compiler diagnostics. 
Their evaluation showed that YATE increased the compilation success rate by 5.96\% points at the class level and 5.68\% points at the method level, while also producing 39.64\% more passing tests compared to plain LLM generation. 
We also observed in our own trials with ChatGPT that a few manual feedback were often enough to produce correct tests and resolve compile errors. 

Moreover, beyond mitigating compilation issues, prior work~\cite{wang-etal-2025-testeval} has shown that LLMs still struggle to focus on specific branches or code elements and require deeper reasoning about program logic. 
To address similar limitations, Yang et al.~\cite{yang2025telpa} proposed \textsc{TELPA}, a feedback-driven framework that identifies ineffective tests as counter-examples and iteratively refines them through program-analysis–guided prompting to reach hard-to-cover branches. 
Inspired by this idea, a PR-aware test generation framework could adopt a comparable refinement mechanism, where tests that fail to exercise modified or newly added lines in a pull request are treated as counter-examples, guiding the LLM to iteratively produce tests that focus on the behavioral changes introduced by the code update.

Therefore, we believe agentic tools can be beneficial for this task, as they treat LLMs as active developers. 
Recent tools—such as \textsc{SWE-Agent} \cite{li2024sweagent}, \textsc{OpenHands} \cite{wang2025openhands}, \textsc{Cline} \cite{cline2025}, \textsc{Cursor} \cite{cursor2025}, and \textsc{Windsurf} \cite{windsurf2025}—show this approach by iteratively planning, executing, and debugging code within real environments.
In addition, giving the agent a tool to evaluate and filter F2P tests should raise the quality of the generated tests. 
With this tool, the agent can track how many actual F2P tests it has produced. 
We anticipate that this approach will be superior to the methods we tried in this paper and may offer a practical solution to the challenges we face.

A second research direction can be a deeper analysis of how a PR's characteristics impact test generation. 
This analysis is especially beneficial when a tool's performance is low. 
It can help identify which characteristics lead to reliable performance, showing where the tools can be used confidently.
It can also determine where to focus efforts to improve the tool's performance.
We have benefited from this method in our own study design; looking at the number of files modified in a PR helped us in our prompt engineering (Section~\ref{sec:promptEngineering}). 
Our RQ3 also focused on visualizing and quantifying the effect of PR size on the effectiveness of the test generators. Based on these findings, we believe further analysis in this direction would be beneficial. 
We observed during manual evaluations that file size can also impact the effectiveness of the test generators. 
Future work could scientifically study the impact of file size and other parameters, such as method size or the importance of the file that has been touched. 

\section{Threats to Validity}
\label{sec:threats}
% We discuss threats to our study across three main areas: internal, construct, and external validity. 
% We describe the steps we took to mitigate these threats, where feasible.
% Many of these limitations we identified are inherent to our study's design. 
% While these limitations affect the study as a whole, we believe they do not invalidate our main finding. 
% This finding is that for both tools, generating F2P tests that focus on a PR remains a difficult task.

\subsection{Internal Validity}

\paragraphTitle{Baseline configuration sensitivity}
EvoSuite depends on settings like the time budget, random seed, and similar parameters. 
In this study, we tried to choose these settings to balance the time and resources required for execution with the tool's performance (Section~\ref{sec:methodology}). 
Additionally, the stochastic nature of GPT-4o could potentially affect the exact numbers we report in another round of execution. 
However, while changing these factors might alter the specific numbers, it would not change our main finding.

% \paragraphTitle{Fair comparison} 
% EvoSuite cannot directly generate test cases for specific PR changes, so it initially generates tests for the entire class instead (Section~\ref{sec:evosuite}). 
% In contrast, we instructed GPT-4o to only generate tests for the changes (diffs) introduced in the PR. 
% To help with this, we even provided GPT-4o with already existing tests to guide it (Section~\ref{sec:threats}). 
% This difference in approach might give GPT-4o an advantage over EvoSuite. 
% In addition, EvoSuite has access to the whole codebase and can see all other files and imports. 
% In contrast, GPT-4o can only see one source code file (among the modified files) and its corresponding tests because of token limitations. This difference puts GPT-4o at a disadvantage. 
% We acknowledge both of these unfair situations. 
% However, they are a result of the tools' fundamental design.

\paragraphTitle{Training data leakage} 
A potential internal threat is the possibility that some PRs in our dataset were part of GPT-4o's pre-training data, as they were created or merged before the model's training cutoff date. 
However, our results show the model did not perform exceptionally well, even with this possible advantage. 
This suggests its true performance on completely unseen data would likely be even lower. 
Therefore, any data leakage would likely bias the results in GPT-4o's favor. 
This actually strengthens our conclusion that PR-aware test generation remains a hard, open problem for current LLMs and highlights that more exploration is needed.

% \subsection{Construct Validity}
% \paragraphTitle{Definition of success (F2P)}
% Our definition of success, as detailed in Section~\ref{sec:methodology}, is finding an F2P. 
% This means a tool produces at least one test that reliably fails or does not compile on the base commit, but successfully compiles and passes on the head commit.
% We also do not measure how many F2P tests a tool produces or how thorough those tests are.
% One single F2P test is enough to count as a success.
% This is a lenient metric, yet the tools still cannot show an outstanding performance, underscoring the need for better solutions.
% Furthermore, using other success criteria could lead to different results. 
% For example, metrics like exposing any regression (even without this strict F2P behavior) or letting developers rate a test's usefulness would likely produce results that are potentially lower than the numbers we reported. Moreover, alternative metrics---such as coverage differences or mutation scores---could also be used and would likely lead to different interpretations of effectiveness, since each metric captures a distinct aspect of test quality. In this study, we adopt the F2P criterion because it offers a clear behavioral signal between the base and head commits and aligns with widely used evaluation practices in benchmarks like SWE-Bench \cite{jimenez2023swe}, which similarly rely on fail-before / pass-after behavior to assess correctness.

\paragraphTitle{Data Cleaning}
During our data cleaning phase, we found multiple PRs that were unsuitable for this study, but our automatic selection process failed to filter them out. 
We manually removed these identified cases from the dataset (Explained in details in Section~\ref{sec:data-collection}).
We mitigate this threat by carefully evaluating the dataset. 
We analyzed all the error cases and sampled the No-F2P cases to verify they are ineffective (Explained in details in Section~\ref{sec:data-collection}).

% \paragraphTitle{Single-Prompt LLM baseline}
% Our study relies on a single prompt, which reflects a common but limited workflow. 
% We recognize that more advanced techniques with better prompts, retrieval steps, or structured decoding, could improve the GPT-4o’s performance. 
% We did experiment with several prompt tweaks (see Section~\ref{sec:promptEngineering}), but compile errors and weak oracles still occurred often.
% We acknowledge that, in theory, a different prompt might change the results. 
% However, our experiments suggest it is unlikely that a better prompt alone would cause a sudden shift in our findings.
% Since we limited our work to prompt engineering for this study, we consider our single-prompt method a baseline for performance, not an upper limit. We will explore these other, better approaches to improve performance in future work.

% \paragraphTitle{Compilation Errors and Test Isolation}
% In our evaluation, we treat a generated suite as an Error if the pre-PR change version of the project exhibits any compilation error triggered by a test method. We do not isolate test methods into separate classes because EvoSuite relies on shared scaffolding and initialization logic. Splitting tests would require intrusive post-processing and alter tool-native behavior. Since our goal is to evaluate the generators as they are used in practice, we treat any compilation error as a hard failure rather than applying post-processing that could artificially improve results.

\subsection{External Validity}
% \paragraphTitle{Repository Bias (Apache and Java 8)} 
% Our dataset is limited to Apache Software Foundation projects. 
% From this group, due to limitations, we further selected only repositories using Java 8 (Section~\ref{sec:data-collection}.) 
% These limitations mean our findings may not apply to projects outside the Apache organization (like industry projects), which may be more actively developed or follow different practices and standards.
% We acknowledge this threat and consider it an inevitable challenge for any study of this type.

\paragraphTitle{PR granularity bias}  
A potential bias exists, as PRs that modify more code or involve multiple files might have a higher chance of producing at least one F2P test. 
To investigate this, we designed Research RQ3 to specifically evaluate if the number of files touched in a PR affects the performance of the tools (as discussed in Section~\ref{sec:results}).
A potential bias for our RQ3 findings stems from the imbalanced distribution of PRs across the granularity bins. 
We acknowledge this bias, even though we attempted to design the bins to hold a balanced number of PRs. 
The imbalance exists because the number of PRs touching only one file is considerably larger than those in the other bins.
While this mirrors the common practice of keeping PRs concise, it under-represents multi-file, cross-module changes.

\section{Conclusion}
Our study focuses on automated unit test generation for Pull Requests (PRs). 
We introduce a new success metric named F2P, which measures the effectiveness of a test suite in capturing PR-related changes. 
We gathered a dataset of 353 Java 8 PRs to evaluate both EvoSuite and GPT-4o.
We observed that in 64\% of PRs, EvoSuite either generated tests that failed or did not compile on the pre-PR version, or it produced no F2P tests at all---indicating limited sensitivity to the specific behavioral changes introduced by the PR.
For GPT-4o, we found that even though it can focus on the PR's diff, its performance is limited by frequent errors.
In fact, 63\% of the tests it generated resulted in an error caused by hallucination, missing imports, or type and visibility mismatches.
We observed that EvoSuite surpassed GPT-4o in this task by yielding a higher F2P rate. 
However, taken together, these results suggest that PR-aware test generation cannot rely on either traditional search-based methods or the use of single-prompt LLMs.

Overall, we believe LLMs offer a key advantage at the PR scope: they can be instructed to focus on the modified lines, affected methods, and nearby tests, making them a natural fit for generating F2P tests. 
However, as our results show, the high rate of errors hinders their ability and results in lower efficiency compared to EvoSuite.
In response, we believe that a feedback-driven, agent-based workflow can potentially solve this issue. 
Such a workflow would iteratively build and run the code, read compiler and runtime signals, and then repair or retarget test candidates until F2P is achieved.

% Our study frames automated test generation  for PR. 
% We introduce a new success metric named F2P which measure the effectiveness of a test suite in capturing PR related changes. 
% We gathered a dataset of 353 Java 8 PRs to evaluate both EvoSuite and GPT-4o. 
% We observed that in 64\% of PRs, EvoSuite yielded no F2P tests, indicating limited sensitivity to the specific behavioral changes introduced by a PR.
% For GPT-4o, despite its potential to focus on the modified region of code, we find that frequent compilation failures are a major obstacle.
% 63\% of the generated tests result in an error caused by hallucination, missing imports/fixtures, and type or visibility mismatches.
% We have observed that EvoSuite has surpassed GPT-4o in this task in terms of yuild more F2P rate. 
% However, taken together, these results suggest that PR-aware test generation cannot rely on either traditional search-based generation or single-prompt LLM use.

% We believe, at the PR scope, LLMs offer a key advantage: they can be instructed to focus on the modified lines, affected methods, and proximate tests, making them a natural fit for generating F2P tests.
% However, the high ratio of errors hinders their ability and results in lower efficiency compared to EvoSuite.
% In response, we believe that a feedback-driven, agent-based workflow can potentially mitigate this issue that iteratively builds and runs base and head, reads compiler and runtime signals, and repairs or retargets candidates until F2P is achieved. 

\bibliographystyle{IEEEtran}
\bibliography{references}

\end{document}

\usepackage{comment}
\begin{comment}

\documentclass[sigconf,review,anonymous]{acmart}

% --- Submission metadata (safe for review; tweak as needed) ---
\setcopyright{acmlicensed}            % no copyright footers in review
\acmYear{2026}
\copyrightyear{2026}
\acmConference[FORGE'26]{3rd ACM International Conference on AI Foundation Models and Software Engineering}{April 12--13, 2026}{Rio de Janeiro, BRAZIL}
\acmISBN{978-1-4503-XXXX-X/2018/06}
\acmDOI{XXXXXXX.XXXXXXX}
\settopmatter{printacmref=false} % hide ACM Reference Format block in review

% ###########################
% ###########################
% ############## Space modification
% ###########################
\AtBeginDocument{%
  \setlength{\textfloatsep}{2pt plus 2pt minus 2pt}   % space above/below floats at top/bottom
  \setlength{\floatsep}{2pt plus 2pt minus 2pt}       % space between floats
  \setlength{\intextsep}{2pt plus 2pt minus 2pt}      % space above/below in-text floats
  \setlength{\dbltextfloatsep}{2pt plus 2pt minus 2pt}
  \setlength{\dblfloatsep}{2pt plus 2pt minus 2pt}
  % Caption gap (acmart uses 'caption' pkg)
  \captionsetup{skip=2pt}                              % space between figure/table and its caption
}
\makeatletter
\renewcommand\section{\@startsection{section}{1}{\z@}%
  {7pt plus 1pt minus 1pt}% space BEFORE (acmart is larger by default)
  {7pt plus 1pt minus 1pt}% space AFTER
  {\normalfont\Large\bfseries}}
\renewcommand\subsection{\@startsection{subsection}{2}{\z@}%
  {1pt plus 1pt minus 1pt}%
  {1pt plus 1pt minus 1pt}%
  {\normalfont\large\bfseries}}
\renewcommand\subsubsection{\@startsection{subsubsection}{3}{\z@}%
  {1pt plus 1pt minus 1pt}%
  {1pt plus 1pt minus 1pt}%
  {\normalfont\normalsize\bfseries}}
\makeatother
% ##########################

% BibTeX macro (as in ACM samples)
\AtBeginDocument{%
  \providecommand\BibTeX{{%
    Bib\TeX}}}

% --- Packages you were using (compatible with acmart) ---
\usepackage{subfiles}
\usepackage{soul}
\usepackage{xcolor}
\usepackage{graphicx}
\usepackage{svg}
\graphicspath{{./images/}}

\usepackage{listings}
\lstset{basicstyle=\ttfamily\small, columns=fullflexible}

% --- Your macros ---
\newcommand{\TODO}[1]{\sethlcolor{yellow}\hl{(TODO: #1)}}
\newcommand{\llm}{GPT-4o}

% pragraphTitle
\newcommand{\paragraphTitle}[1]{\textbf{\textit{#1.}}}

% If you rely on local styling/macros:
\input{typesetting}

% ================================================================
\begin{document}

\title{PR-Aware Automated Unit Test Generation Challenges and Opportunities}

% ----------------------------------------------------------------
% Author metadata (acmart hides this automatically in anonymous mode)
% Uncomment/keep as-is; it will be suppressed with the 'anonymous' option.
% ----------------------------------------------------------------
\author{Vahid Haratian}
\affiliation{
  \institution{Bilkent University}
  \city{Ankara}
  \country{Turkey}
}
\email{vahid.haratian@bilkent.edu.tr}
\orcid{0009-0001-3048-9586}

\author{Atakan Akar}
\affiliation{
  \institution{Bilkent University}
  \city{Ankara}
  \country{Turkey}
}
\email{atakan.akar@ug.bilkent.edu.tr}

\author{Berk Çakar}
\affiliation{
  \institution{Purdue University}
  \city{West Lafayette, IN}
  \country{USA}
}
\email{bcakar@purdue.edu}

\author{Eray Tüzün}
\affiliation{
  \institution{Bilkent University}
  \city{Ankara}
  \country{Turkey}
}
\email{eraytuzun@cs.bilkent.edu.tr}
\orcid{0000-0002-5550-7816}

% ----------------------------------------------------------------
% Abstract, CCS Concepts (optional for review), Keywords
% ----------------------------------------------------------------
\begin{abstract}
\subfile{sections/abstract}
\end{abstract}

\begin{CCSXML}
<ccs2012>
   <concept>
       <concept_id>10011007.10011074.10011099.10011102.10011103</concept_id>
       <concept_desc>Software and its engineering~Software testing and debugging</concept_desc>
       <concept_significance>500</concept_significance>
       </concept>
   <concept>
       <concept_id>10011007.10011006.10011073</concept_id>
       <concept_desc>Software and its engineering~Software maintenance tools</concept_desc>
       <concept_significance>300</concept_significance>
       </concept>
   <concept>
       <concept_id>10011007.10011006.10011071</concept_id>
       <concept_desc>Software and its engineering~Software configuration management and version control systems</concept_desc>
       <concept_significance>100</concept_significance>
       </concept>
 </ccs2012>
\end{CCSXML}

\ccsdesc[500]{Software and its engineering~Software testing and debugging}
\ccsdesc[300]{Software and its engineering~Software maintenance tools}
\ccsdesc[100]{Software and its engineering~Software configuration management and version control systems}

\keywords{software testing, automated unit test generation, large language models, pull requests, incremental test generation, agentic workflows}

\maketitle

% ----------------------------------------------------------------
% Main content
% ----------------------------------------------------------------
\section{Introduction}
\subfile{sections/introduction}

\section{Background}
\subfile{sections/background}

\section{Methodology}
\label{sec:methodology}

\subfile{sections/methodology}

\section{Results}
\subfile{sections/results}

\section{Discussion}
\subfile{sections/discussion}

% \subsection{Future Work}
% \subfile{sections/futureWork}

\section{Threats to Validity}
\subfile{sections/threats}

\section{Conclusion}
\subfile{sections/conclusion}

% ----------------------------------------------------------------
% References
% ----------------------------------------------------------------
\bibliographystyle{ACM-Reference-Format}
\bibliography{references}

\end{document}